\documentstyle[12pt]{article}
\def\beq{\begin{equation}}
\def\eeq{\end{equation}}
\def\bea{\begin{eqnarray}}
\def\eea{\end{eqnarray}}

\catcode`\@=11
\def\marginnote#1{}
\def\ifmath#1{\relax\ifmmode #1\else $#1$\fi}

\def\bold#1{\setbox0=\hbox{$#1$}%
     \kern-.025em\copy0\kern-\wd0
     \kern.05em\copy0\kern-\wd0
     \kern-.025em\raise.0433em\box0 }

\def\GENITEM#1;#2{\par\vskip6pt \hangafter=0 \hangindent=#1
   \Textindent{$ #2$ }\ignorespaces}

\newcount\hour
\newcount\minute
\newtoks\amorpm
\hour=\time\divide\hour by60
\minute=\time{\multiply\hour by60 \global\advance\minute by-
\hour}
\edef\standardtime{{\ifnum\hour<12 \global\amorpm={am}%
    \else\global\amorpm={pm}\advance\hour by-12 \fi
    \ifnum\hour=0 \hour=12 \fi
    \number\hour:\ifnum\minute<100\fi\number\minute\the\amorpm}}
\edef\militarytime{\number\hour:\ifnum\minute<100\fi\number\minute}
\def\draftlabel#1{{\@bsphack\if@filesw {\let\thepage\relax
  \xdef\@gtempa{\write\@auxout{\string
    \newlabel{#1}{{\@currentlabel}{\thepage}}}}}\@gtempa
    \if@nobreak \ifvmode\nobreak\fi\fi\fi\@esphack}
     \gdef\@eqnlabel{#1}}
\def\@eqnlabel{}
\def\@vacuum{}
\def\draftmarginnote#1{\marginpar{\raggedright\scriptsize\tt#1}}
\def\draft{\oddsidemargin -.5truein
        \def\@oddfoot{\sl preliminary draft \hfil
        \rm\thepage\hfil\sl\today\quad\militarytime}
        \let\@evenfoot\@oddfoot \overfullrule 3pt
        \let\label=\draftlabel
        \let\marginnote=\draftmarginnote

\def\@eqnnum{(\theequation)\rlap{\kern\marginparsep\tt\@eqnlabel}%
\global\let\@eqnlabel\@vacuum}  }
\def\preprint{\twocolumn\sloppy\flushbottom\parindent 1em
        \leftmargini 2em\leftmarginv .5em\leftmarginvi .5em
        \oddsidemargin -.5in    \evensidemargin -.5in
        \columnsep 15mm \footheight 0pt
        \textwidth 250mmin      \topmargin  -.4in
        \headheight 12pt \topskip .4in
        \textheight 175mm
        \footskip 0pt

\def\@oddhead{\thepage\hfil\addtocounter{page}{1}\thepage}
        \let\@evenhead\@oddhead \def\@oddfoot{} \def\@evenfoot{}
}
\def\titlepage{\@restonecolfalse\if@twocolumn\@restonecoltrue\o
necolumn
     \else \newpage \fi \thispagestyle{empty}\c@page\z@
        \def\thefootnote{\fnsymbol{footnote}} }
\def\endtitlepage{\if@restonecol\twocolumn \else  \fi
        \def\thefootnote{\arabic{footnote}}
        \setcounter{footnote}{0}}  %\c@footnote\z@ }
\catcode`@=12
\relax
\def\be{\begin{equation}}
\def\ee{\end{equation}}
\def\bea{\begin{eqnarray}}
\def\eea{\end{eqnarray}}

\def\mst11{m_{\;\widetilde{t}_{1}}}

\def\mst22{m_{\;\widetilde{t}_{2}}}
\def\mst12{m_{\;\widetilde{t}_{1,2}}}

\def\msb11{m_{\;\widetilde{b}_{1}}}
\def\msb22{m_{\;\widetilde{b}_{2}}}
\def\msb12{m_{\;\widetilde{b}_{1,2}}}

\def\mwidetilde2{\widetilde{m}^{2}}

\relax
\begin{document}
\input epsf

\topmargin-2.5cm
%\draft
%\preprint
%
\begin{titlepage}
\begin{flushright}
astro--ph/0107502 \\
\end{flushright}
\vskip 0.1in
\begin{center}
{\large\bf  Adiabatic and Isocurvature Perturbations
from Inflation: }\\
\vskip 0.3cm
{\large\bf Power Spectra and Consistency Relations}

\vskip 2cm
{\bf N. Bartolo}$^{1,2}$,
{ \bf S. Matarrese}$^{1,2}$ 
{ and}
{\bf A. Riotto}$^{2}$

\vskip0.7cm
$^{1}${\it Dipartimento di Fisica di Padova ``G. Galilei''}

\vskip 0.2cm

{\it Via Marzolo 8, Padova I-35131, Italy}

\vskip 0.1cm
       
\begin{center}

{\it and}

\end{center}

\vskip 0.1cm

${^2}${\it INFN, Sezione di Padova}

\vskip 0.2cm

{\it Via Marzolo 8, Padova I-35131, Italy}

\end{center}
\vskip 1cm
\begin{center}
{\bf Abstract}
\end{center}
\begin{quote}

We study adiabatic  and isocurvature
 perturbations produced during a period of cosmological inflation.
 We compute the power spectra 
and cross spectra of the curvature and isocurvature modes, 
as well as the tensor perturbation spectrum in terms of the slow-roll 
parameters. We provide two consistency relations for the  amplitudes 
and spectral indices of the corresponding power  spectra.
These relations represent a definite prediction and a test 
of inflationary models
which 
should be adopted when studying cosmological perturbations  
through the Cosmic Microwave Background
   in forthcoming satellite experiments.

\end{quote}
\vskip1.cm
\begin{flushleft}
July 2001 \\
\end{flushleft}

\end{titlepage}
\setcounter{footnote}{0}
\setcounter{page}{0}
\newpage
%
% BODY
\baselineskip=24pt
\noindent

\section{Introduction}
Inflation is the standard scenario for the generation of 
cosmological perturbations in the universe which are the seeds 
for the large scale structure formation and 
the Cosmic Microwave Background (CMB) anisotropies. 
Many inflationary models have been proposed so far since 
the original proposal  by Guth \cite{Guth}. 
The simplest possibility is to assume the presence of 
a single scalar field $\phi$ 
with a potential $V(\phi)$, undergoing a slow-rolling phase \cite{LythRio}.
The dynamics of the inflationary stage can then be studied  
introducing a set of slow-roll parameters
\cite{Lidseyetal, Liddleetal} which are obtained from $V(\phi)$ 
and its derivatives $V'$, $V''$, $\cdots$, $V^{(n)}$ 
with respect to the inflaton field $\phi$. The physical observables 
can be expressed in terms of these parameters. The scalar perturbations are 
generally expected to be adiabatic, nearly Gaussian distributed and 
(almost) scale-free (i.e. with power-spectra $\propto k^n$).
Furthermore, the tensor modes (gravitational waves) are Gaussian and 
scale-free. The scalar and tensor spectra can be parametrized as
\begin{equation} \label{singlespectra}
A^{2}_{S}(k)=A^{2}_{S}(k_{0})\left( \frac{k}{k_{0}} \right)^{n_{S}-1}\, ,  
\qquad A^{2}_{T}(k)=A^{2}_{T}(k_{0})\left( \frac{k}{k_{0}} \right)^{n_{T}},
\end{equation} 
where $k_0^{-1}$ is a typical length scale 
probed by CMB experiments. The  main observables are four: 
the two amplitudes and the spectral indices 
$n_{T}$ and $n_{S}$. They can be expressed in terms 
of the slow-roll parameters
\begin{equation}
\epsilon=\frac{m_{Pl}^2}{16\pi} \left (\frac{V'}{V} \right) ^2 
\qquad \eta=\frac{m_{Pl}^2}{8\pi} \frac{V''}{V}\, ,
\end{equation}
(with $\epsilon, \eta \ll1$ during slow-roll) via the relations
$n_{S}=1-6\epsilon+2\eta$, $n_{T}=-2\epsilon$ and 
$A^{2}_{T}/A^{2}_{S}=\epsilon$. For single-field models,
\begin{equation} \label{rel}
n_T=-2\epsilon, \quad \frac{A_T^{2}}{A_S^{2}}=\epsilon \quad  
\Rightarrow \quad n_T=-2 \frac{A_T^{2}}{A_S^{2}}\, .
\end{equation}  
The so called {\it consistency} relation $n_{T}=-2A^{2}_{T}/A^{2}_{S}$ 
reduces the number of independent observables to $n_{S}$, 
the relative amplitude of the two spectra and the scalar perturbation 
amplitude (which might be determined by normalizing to COBE data).\\
Analyses of the observed CMB anisotropies have so far assumed this kind 
of power-spectra as far as  the primordial perturbations 
are concerned (see, for example, \cite{CMBconstr}). One should emphasize, 
however, that the theoretical predictions for the initial cosmological 
perturbations should be  at the same level of accuracy as the 
observations in order to constrain the cosmological parameters 
($\Omega_{tot}, \Omega_{b}, h$, etc.). 
The forthcoming set of data on the  CMB anisotropies provided by 
the MAP \cite{MAP} and Planck \cite{Planck} satellites 
are expected to reduce the errors on the determination of the
cosmological parameters to a few percent \cite{MAPPlanckerrors}. 
This implies that the assumption that inflation has been driven by a 
single scalar field may turn out to be an oversimplification and that 
it would be useful to consider alternative possibilities to the simplest 
single-field models of inflation. For instance, adiabaticity and/or 
Gaussianity may not hold
\cite{isocurvatureinflation,nongaussianinflation,peebles}. 
Isocurvature perturbations can be produced during a period of inflation 
if more than one scalar field is present. It could be the case of 
inflation driven by several scalar fields (the so called ``multiple 
inflation''), or one where inflation is driven by a single scalar field 
(the inflaton), with other scalar fields whose energy densities are 
subdominant, but whose fluctuations must be taken into account too 
\cite{iso}. We will use $\phi$ and $\chi_I$ ($I=1,...,K$) to indicate all 
the scalar fields, keeping in mind that, if the case, $\phi$ plays the role 
of the inflaton, and $\chi_I$ of the extra degrees of freedom. It is likely 
that in the early universe there were several scalar fields; moreover, from 
the particle physics point of view, the presence of different scalar fields 
is quite natural. An example is given by the supergravity and (super)string 
models where there are a large number of the so called moduli fields. 
Another example is the theories of extra-dimensions where an 
infinite tower of spin-0 graviscalar Kaluza-Klein excitations appear
\cite{extrareview}.\\
On the other hand, isocurvature perturbations, once generated 
\emph{during} inflation, could not survive \emph{after} inflation ends 
\cite{isocurvatureinflation, Linde, Starobinsky, Polarstarobinsky}. 
If during reheating all the scalar fields decay into the same species, 
the only remaining perturbations will be of adiabatic type.\\
In the case of adiabatic plus isocurvature fluctuations, an interesting 
issue is the possible \emph{correlation} between the two modes of 
perturbation. In fact, until recently, only independent mixtures of adiabatic 
and isocurvature modes were considered \cite{indep}. In Ref. \cite{LanRiaz} 
the effects of the correlation on the CMB anisotropies and on the mass power 
spectrum has been considered. It has been found that several peculiar imprints on the 
CMB spectrum arise. In that case the correlation has been put {\it by hand} 
as an additional parameter for structure formation at the beginnig of the 
radiation dominated era. In Ref. \cite{langlois}, instead, 
a specific realization of a double inflationary model with two non interacting
scalar fields was studied as an example for the origin of the correlation 
during inflation. A clear formalism was introduced in Ref. \cite{gw} to study 
the adiabatic and the isocurvature modes and their cross correlation in the 
case of several scalar fields interacting through a generic potential 
$V(\phi,\chi_I)$. 
In a previous paper \cite{BMT} we have shown that, in the presence of several 
scalar fields, it is natural to expect a mixing and an oscillation mechanism 
between the fluctuations of the scalar fields $\phi$ and $\chi_I$, in a 
manner similar to neutrino oscillations. This can happen even if the energy 
density of the scalar fields $\chi_I$ is much smaller than the energy density 
of the field $\phi$. 
The correlation between the adiabatic and the isocurvature perturbations can 
be read as a result of this oscillation mechanism.\\
The aim of  this paper is to express the spectra for the 
adiabatic and isocurvature modes and their cross-spectrum 
in terms of the slow-roll parameters. 
We will show that, as for the standard single-field case, the physical
observables are not independent, but there exist specific consistency 
relations which are predicted theoretically. Analyses of the present CMB 
anisotropies data coming from the BOOMERang and MAXIMA-1 experiments 
have been recently made \cite{TrottaetalAm} and used to constrain adiabatic 
and isocurvature perturbations; a study of the impact of isocurvature 
perturbation modes in our ability to accurately constrain cosmological 
parameters with the forthcoming MAP and PLANCK measurements has been made 
in Ref. \cite{BucherMoodleyTurok}. 
However, in all these studies the physical observables ({\it i.e.} 
the different amplitudes and spectral indices) have been considered
as independent parameters. Our findings, instead, indicate that the 
interplay between the cosmological perturbations generated during the 
inflationary epoch imposes some consistency relations among the physical 
obervables which could be tested in the future.
The plan of the paper is as follows. In Section $2$ we briefly recall the 
basic definitions of isocurvature and adiabatic perturbations, and define 
the correlation spectrum. In Section $3$ we discuss the generation of the 
correlation during an inflationary period where two scalar fields are 
present, making an expansion of the solutions in slow-roll parameters. 
In section $4$, we derive the expressions of the spectra soon after inflation 
and from these we calculate the amplitude ratios and the spectral indices to 
give the consistency relations between them. Finally, Section $5$ contains 
some concluding remarks. 
    
\section{Basic definitions}
Let us consider a system composed by $N$ components. These could be the 
$N$ scalar fields during inflation or the different species which are 
present deep in the radiation era after inflation.
 Adiabatic perturbations are 
perturbations in the total energy density of the system, while isocurvature 
(or entropic) perturbations leave the total energy density unperturbed by a 
relative fluctuation between the different components of the system. Thus 
adiabatic perturbations are characterized by a perturbation in the intrinsic 
spatial curvature, while the isocurvature perturbations do not perturb the 
curvature.\\
In order to have isocurvature perturbations it is necessary to have more than 
one component and at least one nonzero entropic perturbation 
$S_{\alpha\beta}$ \cite{KodSasa}:
\beq \label{S}
S_{\alpha \beta}\equiv \frac{\delta_{\alpha}}{1+w_{\alpha}} - 
\frac{\delta_{\beta}}{1+w_{\beta}}
\neq 0,
\eeq 
where $\delta_{\alpha}=\delta
\rho_{\alpha}/\rho_{\alpha}$\, , $w_{\alpha}=p_{\alpha}/\rho_{\alpha}$ 
(the ratio of the pressure to the energy density), and $\alpha$ and $\beta$ 
stand for any two components of the system. 
$S_{\alpha \beta}$ is a gauge-invariant quantity and measures the relative 
fluctuations between the different components.  Adiabatic 
perturbations are characterized by having $S_{\alpha \beta}=0$ for all of 
the components. Thus in general there will be one adiabatic perturbation 
mode and $N-1$ independent isocurvature modes and one must consider 
adiabatic plus isocurvature perturbations.\\
For a generic cosmological perturbation $\Delta(\bf{x})$, it is standard 
to define its dimensionless power spectrum ${\mathcal{P}}_\Delta$ as:
\begin{equation} \label{def1}
\langle {\Delta}_{\bf{k}} {\Delta}_{\bf{k'}} \rangle=2\pi^2\, k^{-3}\, 
{\mathcal{P}}_\Delta(k)\, \delta(\bf{k} + \bf{k'})\, ,
\end{equation}
where the angular brackets denote ensemble averages and ${\Delta}_{\bf{k}}$ 
is the Fourier transform of $\Delta(\bf{x})$: 
\begin{equation}
{\Delta}_{\bf{k}}=\frac{1}{(2\pi)^{3/2}} \int \textnormal{d}^{3}{\bf{x}}\, 
e^{-i{\bf{k}}\cdot{\bf{x}}}\,\Delta(\bf{x})\, . 
\end{equation}  
Thus for two quantities $\Delta_{1}(\bf{x})$ and  $\Delta_{2}(\bf{x})$ it 
can be defined a cross spectrum as: 
\beq
\langle {\Delta}_{1\bf{k}} {\Delta}_{2\bf{k'}} \rangle=2\pi^2\, k^{-3}\, 
{\mathcal{C}}_{\Delta_{1}\Delta_{2}}(k)\, \delta(\bf{k} + \bf{k'}).  
\eeq     
\section{Adiabatic and Isocurvature perturbations from inflation: 
a slow-roll formalism}
As already mentioned in Section $1$, adiabatic and isocurvature perturbations 
can be produced during a period of inflation in which more than one scalar 
field is present. One of the difficulties in studying mixtures of isocurvature
and adiabatic perturbations produced during inflation is that, in general, 
one cannot trace back the adiabatic mode to the perturbations of some of 
these scalar fields only, and the entropic modes to the perturbations of 
the remaining scalar fields. Rather the fluctuations of all of the scalar 
fields contribute to the adiabatic and isocurvature modes. On the other 
hand, this is the reason why one must expect a correlation between them. 
In this respect the authors of Ref. \cite{gw} have provided a general 
formalism to better disentangling the adiabatic and isocurvature 
perturbation modes.\\
Let us now enter into the details. For simplicity we will restrict here to 
the case of two fields, $\phi$ and $\chi$ with a generic potential 
$V(\phi,\chi)$.
In order to study the field perturbations $\delta \phi$ and $\delta \chi$, 
we can write the line element for scalar perturbations of the metric as:
\begin{eqnarray}
ds^2 &=& - (1+2A)dt^2 + 2aB_{,i}dx^idt \nonumber\\ &+&
a^2\left[ (1-2\psi)\delta_{ij} + 2E_{,ij}\right] dx^idx^j.
\end{eqnarray}  
Thus the equation for the evolution of the perturbation 
$\delta \phi_I$ ($I=1,2$ and 
$\delta \phi_1=\delta \phi$, $\delta \phi_2=\delta \chi$)
with comoving wavenumber $k=2\pi a/\lambda$  for a mode with physical 
wavelength $\lambda$ is:
\begin{eqnarray}
&& \ddot{\delta\phi}_I + 3H\dot{\delta\phi}_I
+ \frac{k^2}{a^2} \delta\phi_I + \sum_J V_{\phi_I\phi_J}
\delta\phi_J
\nonumber\\ &=&
-2V_{\phi_I}A + \dot\phi_I \left[ \dot{A} + 3\dot{\psi} +
\frac{k^2}{a^2} (a^2\dot{E}-aB) \right],
\end{eqnarray}
where the dots stand for time derivatives.\\
In the following we will recall the basic equations and results of Ref. 
\cite{gw}.
It is possible to define the adiabatic and entropy fields 
($\delta A$ and $\delta s$ respectively) in terms of the original ones
$\delta \phi$, $\delta \chi$ as:
\beq
\delta A=(\cos \beta)\delta \phi+(\sin \beta)\delta \chi
\eeq 
and
\beq
\delta s=(\cos \beta)\delta \chi-(\sin \beta)\delta \phi,
\eeq
where 
\beq \label{angledef}
\cos \beta=\frac{\dot{\phi}}{\sqrt{\dot{\phi}^{2}+\dot{\chi}^{2}}}\, , \qquad  
\sin \beta=\frac{\dot{\chi}}{\sqrt{\dot{\phi}^{2}+\dot{\chi}^{2}}}\, .
\eeq 
Introducing the gauge-invariant Sasaki-Mukhanov variables \cite{var}
\beq
Q_I \equiv \delta \phi_I+\frac{\dot{\phi_I}}{H} \psi,
\eeq
it can be checked that $\delta A$ and $\delta s$ can be rewritten as
\beq \label{dA}
Q_A=(\cos \beta)Q_\phi+(\sin \beta)Q_\chi, \label{A}
\eeq
\beq \label{ds}
\delta s=(\cos \beta)Q_\chi-(\sin \beta)Q_\phi \label{s}.
\eeq
Note that the entropy field $\delta s$ is gauge-invariant.\\
The curvature perturbation \cite{CURV}
\beq
\label{def:calR} 
{\cal R}= H  \sum_I \left(\frac{\dot\varphi_I}{\sum_{J=1}^{N} 
\dot\varphi_J^2} \right) Q_{I} 
\eeq
can be written in terms of the adiabatic field $Q_A$ in the same way 
as in the single scalar field case:
\beq \label{R}
{\mathcal{R}}= \frac{H}{\dot{A}} Q_A \, .
\eeq
The master equations are the evolution equations for the quantites defined 
in Eqs. (\ref{A}) and (\ref{s}). They read:
\beq \label{entropy}
\ddot{\delta s} + 3H\dot{\delta s} + \left(\frac{k^2}{a^2}
+ V_{ss} + 3\dot{\beta}^2 \right) \delta s =
{\dot\beta\over\dot{A}} {k^2 \over 2\pi G a^2} \Psi\,
\eeq
and
\begin{eqnarray}
\label{adiabatic}
&& \ddot{Q}_A + 3 H \dot{Q}_A +
\left[ \frac{k^2}{a^2} + V_{AA} - \dot{\beta}^2 
- {8\pi G\over a^3} \left( {a^3\dot{A}^2\over H} 
\right)^{\displaystyle{\cdot}} \right] Q_A \nonumber \\
&&~~~{}= 2(\dot\beta\delta s)^{\displaystyle{\cdot}} - 2\left(
{V_A \over \dot{A}} + {\dot{H}\over H} \right)
\dot\beta\delta s \,,
\end{eqnarray}
where $V_{ss}=(\sin ^2 \beta)V_{\phi\phi}-(\sin 2\beta)V_{\phi\chi}+(\cos^2 
\beta)V_{\chi\chi}$, $\dot{A}=(\cos \beta)\dot \phi+(\sin \beta)\dot{\chi}$, 
\linebreak $V_{AA}=(\sin ^2 \beta)V_{\chi\chi}
+(\sin 2\beta)V_{\phi\chi}+(\cos^2 \beta)V_{\phi\phi}$, 
$V_{A}=(\cos \beta)V_\phi+(\sin \beta)V_\chi$; $\psi=\Psi$ in the 
longitudinal gauge, and we use the notation 
$V_{\phi_I}=\partial V/\partial \phi_I$.\\
Following Ref. \cite{gw}, let us take at horizon crossing during inflation:
\beq \label{init}
Q_I \arrowvert_{k=aH} \approx \frac{H_k}{\sqrt{2k^3}}e_I({\bf k}),
\eeq 
where $I=\phi,\chi$, $H_k$ is the Hubble parameter when the mode crosses the 
horizon ({\it i.e} $a_kH_k=k$) 
and $e_\phi$ and $e_\chi$ are independent random 
variables satisfying
\begin{equation}
\label{eq:Gaussian} 
\langle e_I({\bf k})\rangle = 0 \,, \quad \langle
e_I({\bf k})e_J^*({\bf k}')\rangle = \delta_{IJ}\, \delta({\bf k}-{\bf k}')\, .
\end{equation}
These initial conditions
are strictly valid only for modes well within the horizon. Indeed,
as emphasized in Ref. \cite{BMT}, curvature and isocorvature
perturbations become cross-correlated as soon as
they leave the horizon
when the oscillations between these two modes is resonantly
amplified. 

For super-horizon scales, $k\ll aH$, we can neglect all terms proportional to
$k^2/a^2$ in Eqs. (\ref{entropy}) and (\ref{adiabatic}), 
and consider only the non-decreasing modes which amounts to neglecting the 
second time derivatives. Thus it follows: 
\begin{eqnarray}
Q_A &\simeq& A f(t) + P(t) \,, \label{solutionad}
\end{eqnarray}
\begin{eqnarray}
\delta s &\simeq& B g(t)\,,    \label{solutionis}
\end{eqnarray}
where $f(t)$ is the general solution for the homogeneous part of Eq. 
(\ref{adiabatic}), $P(t)$ is a particular integral of the full 
Eq. (\ref{adiabatic}), and $g(t)$ is the general solution of
Eq. (\ref{entropy}). The amplitudes $A(k)$ and $B(k)$ are given by:
\begin{equation}
A({\bf k}) \approx \frac{H_k}{\sqrt{2k^3}} e_A({\bf k})\, \,\, ,
B({\bf k}) \approx  \frac{H_k}{\sqrt{2k^3}} e_s({\bf{k}}) \, ,
\label{amplitudes}
\end{equation}
where $e_A({\bf k})$ 
and $e_s({\bf k})$ are random variables satisfing the same 
relations of Eq. (\ref{eq:Gaussian}). $P(t)$ can be written as 
$P(t)=B\tilde{P}(t)$. From Eqs. (\ref{solutionad}), (\ref{solutionis}) 
and (\ref{amplitudes}) one gets the expression for $Q_A$ and $\delta s$ 
spectra and their cross-correlation during inflation:
\begin{eqnarray} 
{\cal P}_{Q_A} \label{powerad}
&\simeq& \left( {H_k\over2\pi} \right)^2
\left[ |f^2| + |\tilde{P}^2|\right] \,,
\end{eqnarray}
\begin{eqnarray} \label{powerentr}
{\cal P}_{\delta s}
&\simeq& \left( {H_k\over2\pi} \right)^2 |g^2| \, ,
\end{eqnarray}
\begin{eqnarray} \label{powercross} 
{\mathcal{C}}_{Q_A\delta s}  &\simeq&  \left( {H_k\over2\pi} \right)^2 
\, g\tilde{P}.   
\end{eqnarray}
\subsection{Slow-roll expansion}
The most important comment on the previous formulas is that the correlation 
is nonzero when $\tilde{P}$ is nonzero (we are considering that, in general, 
in a multicomponent system $\delta s \neq 0$). On the other hand $\tilde{P}$ 
is nonzero only when the source term on the right hand-side of 
Eq. (\ref{adiabatic}) is nonzero. This happens when the time derivative 
of the angle $\beta$, defined in Eq. (\ref{angledef}), is not vanishing. 
Note that this is also the condition for the evolution of $Q_A$ and 
$\delta s$ not to be independent, since in this case $\delta s$ 
\emph{feeds} the adiabatic part of perturbations on large scales, 
as observed in Ref. \cite{gw}. In the language of ref. \cite{BMT}
this can be rephrased saying that the probability of oscillation between the 
perturbations of the scalar fields is resonantly
amplified when perturbations cross the horizon and 
the perturbations in the inflaton field may disappear 
at horizon crossing giving rise to perturbations
in scalar fields other than the inflaton. 
Adiabatic and isocurvature perturbations are therefore 
inevitably correlated at the end of inflation.
Provided that $\delta s \neq 0$ -  we can conclude that the correlation will 
be present under the condition $\dot{\beta} \neq 0$. It is remarkable that
only in some special cases this condition is not satisfied. As can be 
checked from Eq. (\ref{angledef}), $\beta$ is \emph{exactly} constant in 
time if there are attractor-like solutions for the evolution of the two 
fields $\phi$ and $\chi$ of the kind $\dot{\chi} \propto \dot{\phi}$. For 
example, this is the case of the models of assisted inflation 
\cite{assistedinfl}. Therefore, if the entropic modes are not strongly 
supressed during inflation, the correlation between isocurvature and 
adiabatic perturbations is quite natural to arise.\\
Now let us introduce the following generalization of the slow-roll
parameters (see Eq. (2)) in the case of two scalar fields:
\beq
\epsilon_{I} =  \frac{m^{2}_{Pl}}{16\pi} \left( \frac{V_{\phi_I}}{V} 
\right)^2 \quad \textnormal{and} \quad \eta_{IJ}=\frac{m^{2}_{Pl}}{8\pi} 
\frac{V_{\phi_{I}\phi_{J}}}{V}\, ,  
\eeq
where $V_{\phi_I}=\partial V/ \partial \phi_I$, and $\phi_I=\phi$ or $\chi$.\\
We have expanded the master equations (\ref{entropy}) and (\ref{adiabatic}) 
to lowest order in these parameters, since during inflation $\epsilon_{I}$ 
and $ \eta_{IJ}$ are $\ll 1$. In the following we will quote only the main 
results. More technical details can be found in the Appendix {\bf A}.\\
For non decreasing modes and $k \ll aH$ the Eq. (\ref{entropy}) can be 
written as:
\beq \label{entropicapprox}
\dot{\delta s} = -\frac{1}{3H} \left(V_{ss} + 3\dot{\beta}^2 \right)  
\delta s. 
\eeq  
Note that $\mu_s ^2 \equiv V_{ss} + 3\dot{\beta}^2$ is the effective mass for 
the entropy field. To lowest order it is given by: 
\beq \label{entropysr}
- \frac{\mu_s ^2}{3H^2} = - \frac{\epsilon_\chi}{\epsilon_{tot}} 
\eta_{\phi\phi}+2 \frac{(\pm \sqrt{\epsilon_\phi})(\pm 
\sqrt{\epsilon_\chi})}{\epsilon_{tot}} \eta_{\phi\chi} - 
\frac{\epsilon_\phi}{\epsilon_{tot}} \eta_{\chi\chi}\, .
\eeq
The sign $\pm$ stands for the cases $\dot{\phi} \, (\dot{\chi}) \, >0  \,  
\textnormal{and} \, <0$ respectively, and $\epsilon_{tot}$ stands for
$(\epsilon_\phi+\epsilon_\chi)$. \\ 
Starting from Eq. (\ref{solutionis}) 
the resulting solution for $\delta s$ will be:
\begin{eqnarray} \label{solutionissr}
\delta s & \simeq & B(k)g(t) \nonumber  
\\ & =  & B(k) \exp\left[ \left( - \frac{\epsilon_\chi}{\epsilon_{tot}} 
\eta_{\phi\phi}+2 \frac{(\pm \sqrt{\epsilon_\phi}) 
(\pm \sqrt{\epsilon_\chi})}{\epsilon_{tot}} \eta_{\phi\chi} 
- \frac{\epsilon_\phi}{\epsilon_{tot}} \eta_{\chi\chi} \right) 
(N_k-N(t)) \right] \nonumber \, ,
\\ 
\end{eqnarray}
where $N_k-N(t)=\int _{t_k}^{t} Hdt$. $N(t)=\int _{t}^{t_f} Hdt$, with 
$t_f$ the time inflation ends, is the number of e-folds between the end of 
inflation and a generic instant $t$, $N_k=\int _{t_k}^{t_f} Hdt = 
\ln(a_f/a_k)$ is the number of e-folds between the time $t_k$ the mode 
crosses the horizon and the end of inflation. Typically, $N_k\simeq 60$
as far as large scale CMB anisotropies are concerned.\\
In order to write Eq. 
(\ref{solutionissr}), we have neglected the time dependence of the term that 
appears as a combination of the slow-roll parameters, since its time 
derivative is ${\mathcal{O}}(\epsilon^2,\eta^2)$  
\footnote{With ${\cal{O}}(\epsilon, \eta)$ and 
${\cal{O}}(\epsilon^2, \eta^2)$ we indicate general combinations of the 
slow-roll parameters of lowest order or next order respectively.} 
and so we have extracted this term out of the integral $N_k-N(t)$. 
Since it can be 
treated as a constant, it can be evaluated at horizon crossing, $k=aH$. 
At the end of inflation $\delta s$ will be:
\beq \label{solution is sr end} 
\delta s\arrowvert_{t_f} =  B(k) \exp\left[ \left( - 
\frac{\epsilon_\chi}{\epsilon_{tot}}
\eta_{\phi\phi}+2 \frac{(\pm \sqrt{\epsilon_\phi})
(\pm \sqrt{\epsilon_\chi})}{\epsilon_{tot}} \eta_{\phi\chi} - 
\frac{\epsilon_\phi}{\epsilon_{tot}} \eta_{\chi\chi} \right) 
N_k \right].
\eeq
As for the adiabatic mode, Eq. (\ref{adiabatic}) can be written as:
\beq \label{adiabaticapprox}
\dot{Q}_A = - \frac{1}{3H}
\left[V_{AA} - \dot{\beta}^2 - {8\pi
G\over a^3} \left( {a^3\dot{A}^2\over H}
\right)^{\displaystyle{\cdot}} \right] Q_A \nonumber \\
+ \frac{2}{3H} \left[ (\dot\beta\delta s)^{\displaystyle{\cdot}} - \left(
{V_A \over \dot{A}} + {\dot{H}\over H} \right) 
\dot\beta\delta s \right].
\eeq
Putting the entropic solution (\ref{solutionissr}) into Eq. 
(\ref{adiabaticapprox}), and following the same procedure of expansion 
in the slow-roll parameters,  we find the adiabatic solution 
(\ref{solutionad}):
\beq \label{solutionad1}
f(t) \arrowvert_{t_f} = \exp \left[ \left(- 
\frac{\epsilon_\chi}{\epsilon_{tot}}\eta_{\chi \chi}- 
\frac{\epsilon_\phi}{\epsilon_{tot}} \eta_{\phi\phi}-2 
\frac{(\pm \sqrt{\epsilon_\phi})(\pm \sqrt{\epsilon_\chi})}
{\epsilon_{tot}} \eta_{\phi\chi}+2\epsilon_{tot} 
\right) N_k \right]
\eeq
and
\beq \label{solutionad2} 
\tilde{P}(t) \arrowvert_{t_f} = 2 \left[ \frac{\dot{\beta}}{H} 
\right]_{l.o.}\, g(t) \arrowvert_{t_f}\, \, \frac{1}{C}(e^{CN_k}-1)\, ,
\eeq 
where $\left[ \frac{\dot{\beta}}{H} \right]_{l.o.}$ is the expression 
of $\dot{\beta}/H$ to lowest order:
\begin{equation}
\left[ \frac{\dot{\beta}}{H} \right]_{l.o.} = \frac
{\epsilon_\phi-\epsilon_\chi}{\epsilon_{tot}}\eta_{\phi\chi}+ 
\frac{(\pm \sqrt{\epsilon_\phi})(\pm \sqrt{\epsilon_\chi})}{\epsilon_{tot}}
(\eta_{\phi\phi}-\eta_{\chi\chi})
\end{equation}
and $C$ is given by:
\beq \label{Cdef}
C = \frac{\epsilon_\phi-\epsilon_\chi}{\epsilon_{tot}}
\eta_{\chi\chi}+\frac{\epsilon_\chi-\epsilon_\phi}{\epsilon_{tot}}
\eta_{\phi\phi}-4\, \frac{(\pm \sqrt{\epsilon_\phi})
(\pm \sqrt{\epsilon_\chi})}{\epsilon_{tot}}\eta_{\phi\chi} 
+2 \epsilon_{tot}\, .
\eeq
Again $\left[ \frac{\dot{\beta}}{H} \right]_{l.o.}$ and $C$ - 
which are ${\mathcal{O}}(\epsilon, \eta)$ - have been treated as constant 
and can be taken at horizon crossing when $k=aH$.\\
Now we are able to give the expressions for the spectra (\ref{powerad}), 
(\ref{powerentr}) and (\ref{powercross}):
\begin{eqnarray} \label{poowerad sr} 
{\cal P}_{Q_A} = 
\left( {H_k\over2\pi} \right)^2\, \arrowvert f^2(t) 
\arrowvert _{t_f}\, \left[1+4\,\left[ \frac{\dot{\beta}}{H}\ \right]_{l.o.}^2\, 
\frac{1}{C^2}(1-e^{-CN_k})^2 \right]\, ,
\end{eqnarray}
\beq \label{powerentr sr}
{\cal P}_{\delta s} =
\left( {H_k\over2\pi} \right)^2\,  \exp\left[ \left( - 2 \frac{\epsilon_\chi}
{\epsilon_{tot}} \eta_{\phi\phi}+4 \frac{(\pm \sqrt{\epsilon_\phi})
(\pm \sqrt{\epsilon_\chi})}{\epsilon_{tot}} \eta_{\phi\chi} 
- 2  \frac{\epsilon_\phi}{\epsilon_{tot}} \eta_{\chi\chi} \right) 
N_k \right]
\eeq
and
\beq \label{powercross sr}
{\mathcal{C}}_{Q_A\delta s} =  \left( {H_k\over2\pi} \right)^2\, 2 \, \, 
\left[ \frac{\dot{\beta}}{H} \right]_{l.o.}  g^2(t)\arrowvert_{t_f}\, \, \frac{1}{C}
(e^{CN_k}-1)
\eeq
Since the isocurvature perturbation $\delta s$ is a source
for the adiabatic one, the curvature perturbation spectrum, which 
characterizes the adiabatic mode, does not remain constant during inflation 
in general, unlike the single field case (see, for example, Ref. \cite{BW}). 
This is the reason why we have evaluated all the previous  expressions at the 
end of inflation. In the next section we will specify the initial conditions 
in the post inflationary epoch.\\
A few comments are in order here. As can be seen in Eq. (\ref{powercross sr}) 
the cross correlation is proportional to $\dot{\beta}$, as already mentioned 
at the beginnig of this section. Moreover it depends on the factor 
$e^{CN_k}$,  which is the ratio between $f$ and $g$. In other words, $C$ is 
$(\mu_s^2 - \mu_A ^2)/3H^2$ to lowest order, where $\mu_s^2$ and $\mu_A^2$ 
are the effective masses for the entropic and adiabatic perturbations 
(the terms proportional to $\delta s$ and $Q_A$ in Eqs. 
(\ref{entropicapprox}), (\ref{adiabaticapprox})). This means that, in order 
to have a strong correlation, what is important is just the \emph{relative} 
magnitude of the adiabatic 
and the entropic masses, even if they are both ${\cal{O}}(\epsilon)$. 
This result is in agreement with our previous
findings \cite{BMT}, where we have demonstrated that the correlation emerges 
as soon as there is a mixing between the original fields $\phi$ and $\chi$
and that this mixing can be large even if 
the masses of the scalar fields are all  ${\cal{O}}(\epsilon_I, \eta_{IJ})$. 
  
\section{Initial conditions in the post inflationary epoch}
In the following we will assume that the mixing between the scalar fields is 
negligible after inflation and that, for example, the field $\phi$ decays 
into ``ordinary'' 
matter (present-day photons, neutrinos and baryons), and 
the scalar field $\chi$ decays only into cold dark matter. The field 
$\chi$ could also not decay, as it happens in axion models. In fact, if during 
reheating all the scalar fields decay into the same species, the
 perturbations will be only of adiabatic type deep in the radiation era: 
no relative fluctuations $S_{\alpha\beta}$ is generated. In the present 
case a CDM-isocurvature mode will survive after inflation. Using the 
notation of Section $2$ and Ref. \cite{langlois}, we can write:
\beq
\delta_{CDM}=S_{CDM-{\rm rest}}+\delta_{A}\, , \qquad     
\delta_{A}=\frac{3}{4}\delta_{\gamma}=\frac{3}{4}\delta_{\nu}=\delta_{b}
\eeq
where $\delta_{A}$ specifies the amplitude of the adiabatic 
mode of perturbations, and ``rest'' stands for ordinary matter.\\
In order to set the initial conditions for the evolution of  cosmological 
perturbations, and  which can be used in some numerical codes calculating the 
CMB anisotropies, we must link the two relevant quantities $S_{CDM-rest}$ 
and $\cal{R}$ deep in the radiation era to the inflationary 
quantities $\delta s$ and $Q_A$.\\
For the adiabatic perturbations this is immediate from Eq. (\ref{R}):
\beq \label{link1}
{\cal{R}}_{rad}= \frac{H}{\dot{A}} Q_A\,
\eeq
where the right hand-side of this equation is evaluated at the end of 
inflation.\\
As far as $S_{CDM-rest}$ is concerned, it is useful to introduce the 
following quantity:
\beq
\delta_{\chi \phi} \equiv \frac{\delta \chi}{\dot{\chi}}-\frac{\delta \phi}
{\dot{\phi}}\, .
\eeq
For the scalar fields $\phi$ and $\chi$ the isocurvature perturbation 
$S_{\chi\phi}$, Eq. (\ref{S}), results $S_{\chi \phi}=a^3 
d{(\delta_{\chi \phi}/a^3)}/dt$ \cite{HwangNoh}.\\
On the other hand:
\beq
\delta s =\frac{\dot{\chi}\ \dot{\phi}}{\sqrt{\dot{\chi}^2+\dot{\phi}^2}}\ 
\delta_{\chi \phi} \, .
\eeq 
Then, to lowest order in the slow-roll parameters, one finds:
\beq \label{link2}
S_{\chi \phi}=-3\ \frac{\sqrt{4\pi}}{m_{Pl}}\ \frac{\sqrt{\epsilon_{tot}}}
{(\pm \sqrt{\epsilon_\phi})(\pm \sqrt{\epsilon_\chi})} \delta s \, .
\eeq 
To match to the radiation epoch we take $S_{CDM-{\rm rest}}=S_{\chi \phi}$ 
at the end of inflation.\\
\subsection{Observables: Amplitudes and  spectral indices}
In this subsection we will give the explicit expressions for the power spectra 
of the adiabatic and isocurvature modes, and their cross correlation. 
To lowest order, they can be written as power laws $\propto k^n$, in a way 
analogous to the single scalar field models (cfr. Eq. (1)). 
This means that there will be three amplitudes and
 three spectral indices. Moreover we have taken into account also the 
tensor perturbation (gravitational-wave) spectrum, yielding a total 
of four amplitudes and four spectral indices. Indeed, we must consider 
the normalization
 that fixes one amplitude and will bring to \emph{seven} 
observables.\\
On the other hand the reader should 
remind that we have introduced \emph{five} slow-roll parameters. In the 
single field case there are three observables (the tensor to scalar amplitude 
ratio $A^2_{T}/A^2_S$, $n_S$ and $n_T$), and one finds one consistency 
relation between  $A^2_{T}/A^2_S$ and $n_T$ (see Eq. (\ref{rel})). Thus in 
the present case we expect to find \emph{two} consistency relations between 
the observables.  
To fit the CMB anisotropies one must consider the initial fluctuation spectra
with their amplitudes and spectral indices. The existence of such consistency 
relations means that not all the amplitudes and spectral indices must be 
considered as independent.\\
For the curvature perturbation ${\cal{R}}$, it results from 
Eqs. (\ref{powerad}) and(\ref{link1}):
\beq \label{spectrumR}
{\cal{P}}_{\cal{R}} = 
\frac{4\pi}{m_{Pl}^2}\left. \left( \frac{H_k}{2\pi} \right)^2 
\frac{1}{\epsilon_{tot}} \left[ |f^2| + |\tilde{P}^2| \right]\right|_{t_f}\, ,
\eeq 
where we have used the fact that $(\dot{A}/H)^2=m_{Pl}^2/4\pi\, \, 
\epsilon_{tot}$.\\
If not written otherwise, we intend this and all the subsequent expressions 
evaluated at the end of inflation for the reasons explained at the end of 
Section $3.1$.\\  
For the isocurvature perturbation $S$, we can write from 
Eqs. (\ref{powerentr}) and (\ref{link2}):
\beq \label{Sspectrum}
{\cal{P}}_{S} = 9\,\left. \frac{4\pi}{m^2_{Pl}}\, \left( \frac{H_{k}}{2\pi} \right)^2 
\frac{\epsilon_{tot}}{\epsilon_\phi\epsilon_\chi}\, 
\arrowvert g^2 \arrowvert\right|_{t_f}\, .
\eeq
Finally, for the cross-spectrum ${\cal{P_{\cal C}}}$ we find  from Eqs. 
(\ref{powercross}) and (\ref{powercross sr}):
\beq \label{crossspectrum}
{\cal{P}}_{\cal{C}} = -6\left.\, \frac{4\pi}{m_{Pl}^2}\, 
\left[ \frac{\dot{\beta}}{H} \right]_{l.o.}\, \left( \frac{H_{k}}{2\pi} \right)^2\, 
\frac{1}{C}\, (e^{CN_k}-1)\,  \frac{1}{(\pm \sqrt{\epsilon_\phi})
(\pm \sqrt{\epsilon_\chi})}\, g^2\right|_{t_f}\, .
\eeq
Now we can calculate the spectral indices to lowest order. 
They are defined as \footnote{Note that 
the correlation can be positive or negative. 
In this case the spectral index can be defined as $n-1 \equiv 
\frac{d \ln |{\cal{P}}|}{d \ln k}$. In the expressions below we have already 
taken it into account.}:
\beq
n-1 \equiv \frac{d \ln {\cal{P}}}{d \ln k}
\eeq
The dependence of the above expressions on the comoving wavenumber $k$ comes 
from  $H_k$, $N_k$, and  those slow-roll parameters which are evaluated at 
horizon crossing, and which are contained in $f$, $g$, 
$[\dot{\beta}/H]_{l.o.}$ and $C$. Therefore, in order to calculate $n$ to 
lowest order, we have made use of the following formula:
\beq
\frac{d \ln {\cal{P}}}{d \ln k} = \frac{d \ln {\cal{P}}}{d \ln (aH)}
{\bigg \arrowvert_{aH=k}} = (1+\epsilon_{tot})\, 
\frac{d \ln {\cal{P}}}{d \ln a}{\bigg \arrowvert_{aH=k}} 
\eeq    
The spectral indices read\footnote{Our definition of the 
isocurvature spectral index differs from $n_{\rm iso}$ as given,
for instance, in refs. 
\cite{peebles}; one has $n_{\rm iso}= n_S-4$.}:  
\begin{eqnarray} \label{indexR}
n_{\cal{R}}-1 &\equiv &
 \frac{d \ln {\cal{P_{\cal{R}}}}}{d \ln k} =  -
6\, \epsilon_{tot}+ 2\, \frac{\epsilon_\chi}{\epsilon_{tot}}\, 
\eta_{\chi\chi}+ 4\, \frac{(\pm \sqrt{\epsilon_\phi}) (\pm \sqrt{\epsilon_\chi})}
{\epsilon_{tot}}\, \eta_{\phi\chi} +2\, \frac{\epsilon_\phi}{\epsilon_{tot}}\, 
\eta_{\phi\phi} \nonumber  \\
& & -8\, \frac{1}{1+\frac{ \arrowvert \tilde{P}^2 \arrowvert}
{\arrowvert f^2 \arrowvert}}\, \left[ \frac{\dot{\beta}}{H} \right]_{l.o.}^2 
\frac{e^{-CN_k}}{C}\, (1-e^{-CN_k})\, , \\
&&\nonumber\\
n_S-1 &\equiv & \frac{d \ln {\cal{P}}_S}{d \ln k}  =   
-2\, \epsilon_{tot}+2\, \frac{\epsilon_\chi}{\epsilon_{tot}}\, 
\eta_{\phi\phi}-4\,\frac{(\pm \sqrt{\epsilon_\phi}) (\pm \sqrt{\epsilon_\chi})}
{\epsilon_{tot}}\, \eta_{\phi\chi}+2\, \frac{\epsilon_\phi}{\epsilon_{tot}}\, 
\eta_{\chi\chi}\, , \\
&&\nonumber\\
n_{\cal{C}}-1 &\equiv&  \frac{d \ln 
\left|{\cal{P}}_{\cal C}\right|}{d \ln k}  =  
n_S-1 -\frac{C}{e^{CN_k}-1}\,  e^{CN_k} \, ,
\end{eqnarray}
where all the slow-roll parameters are evaluated at $k=aH$.\\
As far as the tensor power spectrum is concerned, it is immediate to 
generalize 
the standard result for a single-field model (see, for example, 
\cite{Lidseyetal}). 
To lowest order it is:
\beq
{\cal{P}}_T=\left( \frac{4}{\sqrt{\pi}}\, \frac{H}{m_{Pl}} \right)^2\!  
\bigg \arrowvert_{k=aH}\, \, ,
\eeq
and thus the spectral index $n_T$ reads:
\beq
n_T = \frac{d \ln {\cal{P}}_T}{d \ln k} = -2\, \epsilon_{tot}\, .
\eeq
As can be seen from Eq. (\ref{indexR}), in the case of a single field (for 
example $\phi$) we recover the standard result:
\beq
n_{\cal{R}}-1 = -6\, \epsilon_\phi +2\, \eta_{\phi\phi}.
\eeq
It can be checked that, to lowest order in the slow-roll parameters, these 
spectral indices can be treated as independent of $k$, and so the spectra 
can be approximated, to the desired accuracy, as power laws.

\subsection{Consistency relations}
In order to get the consistency relations we have inverted the equations
defining  the observational 
quantities  ${\cal{P}_R}/{\cal{P}_S}$,  ${\cal{P}_R}/{\cal{P}_C}$, 
${\cal{P}_R}/{\cal{P}}_T$, $n_{\cal{R}}$, 
$n_S$, $n_{\cal{C}}$ and $n_T$ in terms 
of the slow-roll parameters.\\
Let us multiply the power spectra by suitable coefficients 
which are conventional in literature (see Ref. \cite{Lidseyetal}):
$A^2_{\cal{R}} = \frac{4}{25}\, {\cal{P}_R}$, $A^2_{\cal{C}} =\frac{2}{5}\, 
{\cal{P}_C}$ and $A^2_T=\frac{1}{100}\, {\cal{P}}_T$. 

Defining 
\beq
r_T \equiv 
\frac{A^2_T}{A^2_{\cal{R}}} 
\eeq
(not to be confused with the more traditional
tensor-to-scalar quadrupole ratio) and 
\beq
r_{{\cal C}} \equiv   
\frac{A^2_{\cal{C}}} {A_S\, A_{\cal{R}}}\, ,
\eeq
we have found the following 
consistency relations (the interested reader can find the
details in   
Appendix {\bf B})
\begin{eqnarray}
r_T  &=& -\frac{1}{2}\, n_T\, (1-r_{{\cal C}}^2)\, ,
\label{Rel1}\\
&&\nonumber\\      
(n_{\cal{C}}-n_S)r_T & = &-\frac{n_T}{4}(2n_{\cal{C}}-n_S-
n_{\cal{R}}).
\label{Rel2} 
\end{eqnarray}   
A few comments are in order at this point. 
From formula (\ref{Rel1}), one can easily 
recover the single-field model prediction, since in this case 
$A^2_{\cal{C}}$ vanishes and $n_T=-2A^2_T/A^2_{\cal{R}}$. 
We also learn from (\ref{Rel1}) that the tensor to adiabatic 
scalar amplitude ratio is smaller than $-n_T/2$ as soon as  
the adiabatic and entropy modes are cross-correlated. Eq. (\ref{Rel1})  
is a proof of the generic statement that  $r_T\leq -n_T/2$ 
in the multi-component case (see, for example, \cite{PolStar, SasStew}).

Eq. (\ref{Rel2}) applies only when $r_{\cal C}\neq 0$; if the adiabatic and 
isocurvature modes are not correlated (as, for instance, in 
the case of assisted inflation \cite{assistedinfl}) there is only
one consistency relation, which corresponds to the standard formula 
$r_T=-n_T/2$. 

The consistency relation (\ref{Rel2}) can be further simplified if
the slow-roll parameters are smaller than $1/N_k$. In such a case, 
to lowest order we get 
\begin{eqnarray}
r_T  &=& -\frac{1}{2}\, n_T  \, ,
\label{Rel1'}\\
&&\nonumber\\        
n_S & = & n_{\cal{R}}.
\label{Rel2'} 
\end{eqnarray} 

The consistency relations (\ref{Rel1}) and (\ref{Rel2}) (or (\ref{Rel1'}) 
and (\ref{Rel2'})) are the main results of this paper.

\section{Conclusions}
In this paper we have considered the possibility that a CDM-isocurvature 
perturbation mode can survive after an inflationary period in which two 
scalar fields are present. Linking the post inflationary epoch to the 
dynamics of inflation, under the slow-roll conditions, it is possible to 
get the expression for the spectra of the adiabatic, the isocurvature 
modes and their cross-correlation spectrum in terms of the slow-roll 
parameters defined for the two scalar fields. From these expressions two 
consistency relations follow, Eqs. (\ref{Rel1}) and (\ref{Rel2}), 
in analogy to what 
one finds in the single-field case. Thus these relations consitute a strong 
signature of inflation models with more than one scalar field. 
For an analysis of the CMB anisotropy measurements, these relations between 
observables must be taken into account, as a prediction of inflation. 
The main trend is actually to consider all the possible isocurvature 
modes (CDM, baryon, neutrino and neutrino velocity isocurvature modes) in a 
phenomenological way, considering all the amplitudes and the spectral indices 
as independent observables \cite{TrottaetalAm, BucherMoodleyTurok}. 
Even if the 
present treatment does not consider the possible origin of isocurvature 
modes different from the CDM one, it could be easily extended to the case of
more than two scalar fields giving rise to many isocurvature modes. 
Our analysis clearly indicates that, in an inflationary 
scenario for the production of primordial perturbations, not all the 
observables have to be treated as independent. This has strong implications 
for our ability to accurately constrain cosmological parameters 
from CMB measurements in models where both adiabatic and isocurvature 
modes are present. 

%\vskip1cm
%\centerline{\large\bf Acknowledgements}
%\vskip 0.2cm

\vskip 1cm
\section*{Appendix A. Slow-roll expansion}
\setcounter{equation}{0}
\def\theequation{A.\arabic{equation}}
\vskip 0.2cm
Here we report in more detail the calculations leading to 
Eqs. (\ref{entropicapprox}) and (\ref{adiabaticapprox}) to lowest order 
in the slow-roll parameters $\epsilon_I$ and $\eta_{IJ}$.\\
Using  the definiton of the adiabatic and entropic fields, Eqs. (\ref{dA}) and 
(\ref{ds}), and  Eq. (\ref{init}), we obtain  the initial conditions  
at the time $t_k$ of horizon crossing:
\beq \label{initcond}
Q_A \approx \frac{H_k}{\sqrt{2k^3}}e_A({\bf k})\, , \qquad  \delta s  \approx 
\frac{H_k}{\sqrt{2k^3}}e_s({\bf k})\, . 
\eeq
The solution of Eq. (\ref{entropicapprox}) will be:
\beq
\delta s = B(k) \exp\left[ \int_{t_k}^{t} - \frac{\mu^2_s}{3H^2}\, Hdt 
\right]\, ,  
\eeq
where $\mu^2_s=V_{ss}+3\dot{\beta}^2$. Let us recall the explicit expression 
of $V_{ss}$ :
\beq \label{V_ss}
V_{ss}=(\sin ^2 \beta)V_{\phi\phi}-(\sin 2\beta)V_{\phi\chi}+(\cos^2 
\beta)V_{\chi\chi}\, .
\eeq
Using Eq. (\ref{angledef}), we get:
\beq
\sin^2\beta = \frac{\dot{\chi}^2}{\dot{\phi}^2+\dot{\chi}^2} = 
\frac{\epsilon_\chi}{\epsilon_{tot}}\, ,
\eeq 
where $\epsilon_{tot}=\epsilon_\phi+\epsilon_\chi$ and we 
have used the following relations holding to lowest order:
\beq \label{l.o.}
H^2 = \frac{8\pi}{3m_{Pl}^2}V(\phi,\chi)\, \qquad \textnormal{and} \qquad 
\dot{\phi_I} = - \frac{1}{3H} \frac{\partial{V}}{\partial{\phi_I}}\, . 
\eeq
Since $V_{\phi\phi}/H^2 = 3\, \eta_{\phi\phi}$, we obtain:
\beq
\sin^2\beta \,  \frac{V_{\phi\phi}}{H^2} = 3\, \frac{\epsilon_\chi}
{\epsilon_{tot}} \eta_{\phi\phi}\, .
\eeq 
In the same way one calculates the other two terms 
on the right hand-side of Eq. (\ref{V_ss}), leading to (\ref{entropysr}).\\
The quantity $\dot{\beta}^2/3H^2$ may be  neglected since it is 
$O(\epsilon^2, \eta^2)$:
\beq
\frac{\dot{\beta}}{H} = \cos^2 \beta\, \frac{1}{H} \frac{d (\tan \beta)}{dt}
=\frac{\epsilon_\phi} {\epsilon_{tot}} \frac{1}{H}
\frac{d(\tan \beta)}{dt}\, ,
\eeq
with
\begin{eqnarray}
\frac{1}{H} \frac{d (\tan \beta)}{dt} = \frac{1}{H} 
\frac{\ddot{\chi}\dot{\phi}-\dot{\chi}\ddot{\phi}}{\dot{\phi}^2}
& = &\frac{1}{\epsilon_\phi} [- \eta_{\chi\chi}(\pm \sqrt{\epsilon_\chi})
-\eta_{\phi\chi}(\pm \sqrt{\epsilon_\phi})+
\epsilon_{tot}(\pm \sqrt{\epsilon_\chi}) ](\pm \sqrt{\epsilon_\phi})+
\nonumber\\ 
& & \frac{1}{\epsilon_\phi} [\eta_{\phi\phi}(\pm \sqrt{\epsilon_\phi})
+\eta_{\phi\chi}(\pm \sqrt{\epsilon_\chi})-
\epsilon_{tot}(\pm \sqrt{\epsilon_\phi}) ](\pm \sqrt{\epsilon_\chi})\, , 
\nonumber \\
\end{eqnarray}
and thus 
\beq
\left[ \frac{\dot{\beta}}{H} \right]_{l.o.} = \frac{1}{\epsilon_{tot}} 
[(\epsilon_\chi-\epsilon_\phi)\eta_{\phi\chi}+
(\eta_{\phi\phi}-\eta_{\chi\chi})(\pm \sqrt{\epsilon_\chi}) 
(\pm \sqrt{\epsilon_\phi}) ]\, .
\eeq
The function $g(t)$ is:
\beq
g(t) = \exp \left[ \left(- \frac{\epsilon_\chi}{\epsilon_{tot}}\eta_{\phi\phi}+
2 \frac{(\pm \sqrt{\epsilon_\phi})(\pm \sqrt{\epsilon_\chi})}{\epsilon_{tot}}
\eta_{\phi\chi}-\frac{\epsilon_{\phi}}{\epsilon_{tot}}\eta_{\chi\chi} \right)
(N_k-N(t)) 
\right]
\eeq
Recall now  Eq. (\ref{adiabaticapprox}) for the adiabatic perturbation 
$Q_A$:
\beq
\dot{Q_A} = a(t)\, Q_A+ b(t)\, ,
\eeq
where 
\beq
a(t) = - \frac{1}{3H} \left[V_{AA} - \dot{\beta}^2 - {8\pi G\over a^3} 
\left( {a^3\dot{A}^2\over H} \right)^{\displaystyle{\cdot}} \right]\, , 
\eeq
and 
\beq
b(t)=\frac{2}{3H} \left[ (\dot\beta\delta s)^{\displaystyle{\cdot}} - \left(
{V_A \over \dot{A}} + {\dot{H}\over H} \right)\dot\beta\delta s \right]\, .
\eeq
The homogenous solution $f(t)$ is given by $\exp [\int_{t_k}^{t} a(s)\, ds ]$. 
We expand $a(s)$ to lowest order. The same procedure for $V_{ss}/H^2$ holds 
for $V_{AA}/H^2$, and $\dot{\beta}^2/H^2$ is again neglected. As far as the 
last term in $a(t)$ is concerned, one has
\beq 
\frac{1}{H^2} {8\pi G\over a^3} 
\left( {a^3\dot{A}^2\over H} \right)^{\displaystyle{\cdot}} =
3\, \frac{8\pi}{m_{Pl}^2} \frac{\dot{A}^2}{H^2}+ \frac{8\pi}{m_{Pl}^2}
\frac{1}{H^2} \left( \frac{\dot{A}^2}{H} \right)^{\displaystyle{\cdot}}\, .
\eeq
Since $\dot{A} = (\cos \beta)\dot{\phi}+( \sin \beta) \dot{\chi}$, it follows:
\beq
 3\, \frac{8\pi}{m_{Pl}^2} \frac{\dot{A}^2}{H^2} = 3\, \frac{8\pi}{m_{Pl}^2} 
\frac{\dot{\phi}^2+\dot{\chi}^2}{H^2} = 6\, \epsilon_{tot}\, ,
\eeq
and
\beq \label{toneglect}
\frac{1}{H^2} \left( \frac{\dot{A}^2}{H} \right)^{\displaystyle{\cdot}} =
- \frac{m_{Pl}^2}{4\pi} \left[ \left( \frac{\dot{H}}{H^2} \right)^2+
\frac{1}{H} \frac{d}{dt} \left( \frac{\dot{H}}{H^2} \right) \right]\, ,
\eeq
where we have used the formula $\dot{A}^2/H^2 = (-4\pi)^{-1} m_{Pl}^2\, 
\dot{H}/H^2$. Since $- \dot{H}/H^2 = \epsilon_{tot}$ to lowest order, the 
term in Eq. (\ref{toneglect}) is negligible (it is easy to verify that 
the time derivative of $\epsilon_{tot}$ is ${\cal{O}}(\epsilon^2, \eta^2)$).\\
Thus $a(t)$ reads:
\beq
a(t) = -\frac{\epsilon_\chi}{\epsilon_{tot}}\eta_{\chi\chi}-
\frac{\epsilon_\phi}{\epsilon_{tot}}\eta_{\phi\phi}-2 
\frac{(\pm \sqrt{\epsilon_\phi})(\pm \sqrt{\epsilon_\chi})}
{\epsilon_{tot}}\eta_{\phi\chi}+2 \epsilon_{tot}
\eeq
and Eq. (\ref{solutionissr}) follows.\\
Finally, we calculate the particular solution $\tilde{P}(t)$  of the full Eq. 
(\ref{adiabaticapprox}). This is given by $\exp[\int_{t_k}^{t}
a(s)\, ds] \int_{t_k}^{t} \exp[ -\int_{t_k}^{\tau} a(s)\, ds ]\, 
b(\tau)\, d\tau\, .$\\
Inserting $\dot{\delta s}$, Eq.(\ref{entropicapprox}) into $b(t)$, we find:
\beq
\frac{2}{3} \left[ \frac{\ddot{\beta}}{H}-\frac{\dot{\beta}}{H}\frac{1}{3H}
(V_{ss}+3\dot{\beta}^2)-\left( \frac{V_{A}}{\dot{A}}+\frac{\dot{H}}{H} 
 \right) \frac{\dot{\beta}}{H} \right] \delta s\, ,
\eeq
where  $V_A = (\cos \beta) V_{\phi}+(\sin \beta)V_{\chi}$.
To lowest order the only term which survives is $\frac{V_{A}}{\dot{A}}\, 
\frac{\dot{\beta}}{H}$, which to lowest order is given 
by $-3H \left[ \frac{\dot{\beta}}{H} \right]_{l.o.}$.\\
Thus $b(t)$ reads:
\beq
b(t) = 2H \left[ \frac{\dot{\beta}}{H} \right]_{l.o.}\, g(t)\, ,
\eeq  
and $\tilde{P}(t)$ at the end of inflation becomes:
\begin{eqnarray}
\tilde{P}(t)|_{t_f} & = & 2 \left[ \frac{\dot{\beta}}{H} \right]_{l.o.}
e^{\int_{t_k}^{t_f} a(s)\, ds}\, \int_{t_k}^{t_f}e^{- \int_{t_k}^{\tau} 
a(\tau)\, d\tau}\, Hg(\tau) d\tau \nonumber\\
& =  &  2 \left[ \frac{\dot{\beta}}{H} \right]_{l.o.}\, g(t)|_{t_f} 
\int_{t_k}^{t_f} e^{CN(t)}H\, dt  
\end{eqnarray}
where $C$ is given in Eq. (\ref{Cdef}) and we have extracted 
from the integral 
$ \left[ \frac{\dot{\beta}}{H} \right]_{l.o.}$ and $g(t)|_{t_f}$, 
since, to lowest order, they can be considered as being constant. The integral 
can be resolved by the change of variables $Hdt = -dN$ and it yields
 $C^{-1}\, (e^{CN_k}-1)$. Thus Eq. (\ref{solutionad2}) follows.
\vskip 1cm
\section*{Appendix B. Consistency relations}
\setcounter{equation}{0}
\def\theequation{B.\arabic{equation}}
\vskip 0.2cm
To calculate the formulae (\ref{Rel1}) and (\ref{Rel2}), we must take into 
account that there are \emph{seven} observables expressed through \emph{five} 
slow-roll parameters at horizon crossing. To invert the equations defining  
${\cal{P}_R}/{\cal{P}}_S$,  ${\cal{P}_R}/{\cal{P}_C}$, 
${\cal{P}_R}/{\cal{P}}_T$, $n_{\cal{R}}$, $n_S$, $n_{\cal{C}}$ and $n_T$, 
we have made a change of variables using five combinations of the slow-roll 
parameters at horizon crossing which are always found in the expressions for 
the observables. They are:
\begin{eqnarray}
\left[ \frac{\dot{\beta}}{H} \right]_{l.o.} & \equiv & x \\
|f|^2|_{t_f} & \equiv & u>0 \\
g^2|_{t_f} & \equiv & r>0 \\
\epsilon_{tot}|_{k=aH} & & 
\end{eqnarray}  
and
\begin{eqnarray} \label{B.6}
\left(2 \frac{\epsilon_\chi}{\epsilon_{tot}}\eta_{\phi\phi}-
4 \frac{(\pm \sqrt{\epsilon_\phi})(\pm \sqrt{\epsilon_\chi})}{\epsilon_{tot}}
\eta_{\phi\chi}+2 \frac{\epsilon_{\phi}}{\epsilon_{tot}}\eta_{\chi\chi} \right)
\, .
\end{eqnarray}  
Note  that in our results for the spectra, Eqs. (\ref{spectrumR}), (\ref{Sspectrum})
and (\ref{crossspectrum}), there appear also two
expressions in the slow-roll parameters evaluated at the end of inflation, 
$t=t_{f}$, and not only slow-roll parameters at $k=aH$. They are 
$\epsilon_{tot}\arrowvert_{t_f}$ and 
$(\pm \left.\sqrt{\epsilon_\phi})(\pm \sqrt{\epsilon_\chi}) \right|_{t_f}$. 
However, we have taken $\epsilon_{tot}$ equal to one at the end of inflation.
 So one can verify that $|f|^2|_{t_f} 
\sim 1/\epsilon_{tot}|_{k=aH}$, and therefore there are \emph{five} 
variables again: $x$, $u$, $r$, the one defined in Eq. (\ref{B.6}), 
plus $(\pm \sqrt{\epsilon_{\phi}})(\pm \sqrt{\epsilon_{\chi}})|_{t_f}
 \equiv s>0$.\\In these new variables we have considered the following 
equations:
\begin{eqnarray}
n_{S}-n_{\cal{C}} & = & \frac{1}{2 N_k} 
\frac{\sqrt{p}}{\sqrt{p}-1} \ln p \label{B.6} \\
n_{\cal{R}}-n_{S} & =  & -\frac{1}{N_k} \ln p 
-\left[ 16\,  x^2 \frac{N_k}{\ln p}\, 
\frac{\sqrt{p}-1}{p} \right] \frac{1}{F(x,p)} \label{B.7} \\
\frac{4}{9}\, {\cal{P}_R}/{\cal{P}}_S & = & \frac{4}{81}\, 
s^2\, p\, \,  F(x,p) \label{B.8}\\
\frac{9}{4}\, {\cal{P}_C}/{\cal{P}_R} & = & \left[ - \frac{1}{9}\, 
\frac{s}{x}\, \frac{1}{2 N_k}\, \ln p\, \frac{p}{\sqrt{p}-1}\, F(x,p) \right]^{-1} 
\label{B.9}\\
\frac{9}{4}\, {\cal{P}}_T/{\cal{P}_R} & = & 36\, \frac{1}{rp}\, \frac{1}{F(x,p)} 
\label{B.10}\\
n_{T} & = & -2\, \frac{1}{rp} \label{B.11}  
\end{eqnarray}
where
\beq
F(x,p) = 1 + 16\times  N_k^2\, x^2\ \frac{1}{(\ln p)^2}\, 
\frac{(\sqrt{p}-1)^2}{p}\, ,
\eeq
with $p \equiv \frac{u}{r} >0$, and we have used the fact that the quantity $C$ 
given in Eq. (\ref{Cdef}) can be written as $C \approx \frac{1}{N_k}\, 
\ln \sqrt{p}$.\\
Using  Eqs. (\ref{B.8}), (\ref{B.9}), (\ref{B.10}) and (\ref{B.11}) 
 one gets the first consistency relation Eq. (\ref{Rel1}) eliminating
 the variables 
$x$ and $r$.\\
Eqs. (\ref{B.7}) and (\ref{B.10}), eliminating $x$, give the 
following equation:
\beq \label{B.13}
\frac{9}{4} {\cal{P}}_T/{\cal{P}_R} = -\frac{n_T}{2}\, \left[ 36\sqrt{p}+36\, 
N_k\, n_S\, 
\frac{(1-\sqrt{p})}{\ln p}-36\, N_k\, n_{\cal{R}}\, \frac{(1-\sqrt{p})}{\ln p} 
\right]
\eeq 
Using  Eqs. (\ref{B.6}) and (\ref{B.13}), one gets the second consistency 
relation (\ref{Rel2}). The procedure is as follows. We have defined $w 
\equiv \ln p$ and $1-\sqrt{p} \equiv z$. Thus the consistency relation 
is found from the equation $e^{w} \equiv (1-z)^2$, once the 
explicit expressions for $w$ and $z$ are obtained. This is straighforward
leading to:
\begin{eqnarray}
1-z & = & \frac{-(n_{\cal{C}}-n_{S})}{4\, n_T \,
\frac{{\cal{P}_R}}{{\cal{P}}_T}} \\ 
w & = & -2 N_k\, (n_{\cal{C}}-n_S)-36\, N_k\, (n_{\cal{C}}-n_S) n_T \,
\frac{4}{9} \frac{{\cal{P}_R}}{{\cal{P}}_T}-36\, N_k\, (n_{\cal{C}}-n_{\cal{R}}) n_T \,
\frac{4}{9}\, \frac{{\cal{P}_R}}{2{\cal{P}}_T}\, . \nonumber \\ 
\end{eqnarray}

From these equations we get Eq. (\ref{Rel1}) and 
\begin{equation}
\ln \left[ \frac{4(n_{\cal{C}}-n_S)r_T}{n_T(n_S+n_{\cal{R}}
-2n_{\cal{C}})} \right] = 
N_k\, \left[ (n_S-n_{\cal{C}})+\frac{n_T(n_S+n_{\cal{R}}-2n_{\cal{C}})}{4r_T}
 \right]\, .
\label{Rel3} 
\end{equation}

In order to make the solution time-independent, 
consistently with our first-order slow-roll expansion, both sides of 
(\ref{Rel3}) have to vanish. Equation (\ref{Rel2}) then follows. 
The relation (\ref{Rel2'}) in the limit 
$\vert C \vert N_k  \ll 1$ can be easily derived by noting that
$n_{\cal C}-n_S=-1/N_k$ and $2 n_{\cal C}-n_S -n_{\cal R}=-2/N_k$.

\end{document}